\documentclass[a4paper]{jpconf}
\usepackage[
backend=biber,
style=vancouver,
sorting=none
]{biblatex}

\usepackage{graphicx}
\usepackage[labelfont=bf]{caption}
\usepackage{multicol}
\usepackage{subcaption}
\usepackage{tikz}
\usepackage{listings}
\usepackage[utf8]{inputenc}
\usepackage[T1]{fontenc}
\usepackage{zlmtt}
\usepackage{fancyvrb}
\usepackage{color}
\usepackage{amsfonts}
\usepackage[frozencache,cachedir=.]{minted}

\newminted{cpp}{gobble=2, numbersep=3pt, escapeinside=||, xleftmargin=5mm, framesep=2mm, frame=leftline, baselinestretch=1, fontsize=\footnotesize, linenos=true}

\bibliography{clad-acat21}

\begin{document}
\title{GPU Accelerated Automatic Differentiation With Clad}

\author{Ioana Ifrim, Vassil Vassilev and David J Lange}

\address{Department of Physics, Princeton University, Princeton, NJ 08544, USA }

\ead{ii3193@princeton.edu, vassil.vassilev@cern.ch, david.lange@princeton.edu}

\begin{abstract}

Automatic Differentiation (AD) is instrumental for science and industry. It is a tool to evaluate the derivative of a function specified through a computer program. The range of AD application domain spans from Machine Learning to Robotics to High Energy Physics. Computing gradients with the help of AD is guaranteed to be more precise than the numerical alternative and have a low, constant factor more arithmetical operations compared to the original function. Moreover, AD applications to domain problems typically are computationally bound. They are often limited by the computational requirements of high-dimensional parameters and thus can benefit from parallel implementations on graphics processing units (GPUs).

Clad aims to enable differential analysis for C/C++ and CUDA and is a compiler-assisted AD tool available both as a compiler extension and in ROOT. Moreover, Clad works as a plugin extending the Clang compiler; as a plugin extending the interactive interpreter Cling; and as a Jupyter kernel extension based on xeus-cling.

We demonstrate the advantages of parallel gradient computations on GPUs with Clad. We explain how to bring forth a new layer of optimization and a proportional speed up by extending Clad to support CUDA. The gradients of well-behaved C++ functions can be automatically executed on a GPU. The library can be easily integrated into existing frameworks or used interactively. Furthermore, we demonstrate the achieved application performance improvements, including ($\approx$10x) in ROOT histogram fitting and corresponding performance gains from offloading to GPUs.
\end{abstract}

\section{Introduction}

%\textbf{ * What is AD [a paragraph]*}
Many tasks in science and industry are or can be amenable to gradient-based optimization. Gradient-based optimization is an effective way to find optimal parametrization for multiparameter processes that can be expressed with an objective function. The optimization algorithms rely on a function's derivatives or gradients. Multiple approaches to obtain gradients of mathematical functions exist, including manual, numerical, symbolic and automatic/algorithmic. Automatic/algorithmic differentiation (AD) is a computer program transformation that relies on the fact that every program can be decomposed into elementary operations to which the chain rule can be applied. Unlike numeric or symbolic differentiation, AD enables computation of exact gradients and is free of systematic and round-off errors. In addition, AD provides highly efficient means to control the computational complexity of the produced gradients and derivatives~\cite{verma2000introduction}. 

%Its implementation in computer programs can be either through numerical differentiation - finite differences approximations, symbolic differentiation - expression manipulation used in computer algebra systems, manual implementation - of the derivative itself or through Automatic Differentiation (AD). AD is free of systematic and round-off errors due to the fact that the computations are done to machine precision, which for HEP underlines it’s central role in the development of error propagation algorithms and overall algorithmic sensitivity analysis. AD evaluates a derivative of a function using a set of rules which are not influenced by the complexity of a program, given that numerical computations are thought of as compositions of a finite set of elementary operations for which derivatives are known~\cite{verma2000introduction}. 

%\textbf{  * Why AD is important for HEP [a paragraph] * Why AD is important for ML [a paragraph]* }

AD is not new, and has become increasingly popular due to the \textit{backpropagation} technique in machine learning (ML). ML models are trained using large data sets, with an optimization procedure relying on the computation of derivatives for error correction. The scalable AD-produced gradients combined with recent increases in computational capabilities are an enabling factor for sciences such as oceanology and geosciences~\cite{qin2008development}. High Energy Physics (HEP) is well positioned to benefit from gradient-based optimizations allowing for better parameter fitting, sensitivity analysis, Monte Carlo simulations and general statistical analysis for uncertainty propagation through parameter tracking~\cite{ramos2020automatic}.
Data-intensive fields such as ML and HEP advance their scalability by employing various parallelization techniques and hardware. For ML, it is common that part, if not all, of the computations are run on GPGPU-accelerators, taking advantage of parallel computing platforms and programming models, such as CUDA. HEP has invested in systems for GPGPUs for real-time data processing~\cite{vom2020real} and clustering algorithms~\cite{chen2020gpu}. It has also started to reorganize its data analysis and statistical software to be more susceptible towards parallelism. Packages such as RooFit~\cite{hageboeck2020new} can see large speed-ups via gradient-based optimization developments and hardware acceleration support.

The AD program transformation has two distinct modes:  \textit{forward} and \textit{reverse} accumulation mode. The forward accumulation mode has a lower computational cost for functions taking fewer inputs and returning more outputs. The reverse accumulation mode is better suited to "reducing" functions that have fewer outputs and more inputs. The reverse accumulation mode is more difficult to implement, but is critical for many use cases as the execution time complexity of the produced gradient is independent of the number of inputs. The implementation of efficient reverse accumulation AD for parallel systems such as CUDA is challenging as it is a non-trivial task to preserve the parallel properties of the original program. 

This paper describes our progress with the CUDA support of the compiler-assisted AD tool, Clad. We demonstrate the integration of Clad with interactive development services such as Jupyter Notebooks~\cite{xeus-cling}, Cling~\cite{Vasilev_2012}, Clang-Repl~\cite{clang-repl-root} and ROOT~\cite{BRUN199781}.
%Moreover, the compiler-based source to source transformation tool, Clad is implemented as a plugin of the Clang C/C++. 

\section{Background}

%\textbf{* Mention some of they key algorithmic guarantees AD gives us (eg. time complexity independent on inputs). [1 para]}

The AD program transformation is particularly useful for gradient-based optimization because it offers upper bound asymptotic complexity computation guarantees. That is, the cost of the reverse accumulation scales linearly as $O(m)$ where m is the number of output variables. Moreover, the cheap gradient principle~\cite{griewank2008evaluating} demonstrates that the "cost of computing the gradient of a scalar-valued function is nearly the same (often within a factor of 5) as that of simply computing the function itself"~\cite{kakade2018provably}. These guarantees make gradient-descent (also known as error backpropagation in ML) optimization computationally feasible in areas such as deep learning.

AD typically starts by building a computational graph, or a directed acyclic graph of mathematical operations applied to an input. There are two approaches to AD, which differ primarily in the amount of work done before the program execution. The \textit{tracing} approaches construct and process a computational graph for the derivative, by exploiting the ability to overload operators in C++. They perform an evaluation at the time of execution, for every function invocation. \textit{Source transformation} approaches generate a derivative function at compile time to create and optimize the derivative only once. The tracing approach, which is implemented in C++ by ADOL-C\cite{griewank1996algorithm}, CppAD~\cite{bell2011cppad}, and Adept~\cite{hogan2014fast}, is easier to implement and adopt into existing codebases. Source transformation shows better performance, but it usually covers a subset of the language. In C++, Tapenade  ~\cite{hascoet2013tapenade} has a custom language parsing infrastructure.

Being a language of choice for Machine Learning, Python hosts one of the most sophisticated AD systems. \textit{JAX} is a trace-based AD system that differentiates a sub-language of Python oriented towards ML applications~\cite{frostig2018compiling}. \textit{Dex} aims to give better AD asymptotic and parallelism guarantees than JAX for loops with indexing~\cite{paszke2021getting}. Trace-based ML AD tools are developed to exploit ML workflow characteristics such as having little branching, no branching depending on input data or probabilistic draws at runtime, and almost no adaptive algorithms. However, investigations into further generalized approaches to AD have shown superior performance~\cite{RFC:C++Gradients}.

HEP projects that exploit AD include work on MadJAX~\cite{madjax2021ACAT}, ACTS~\cite{ai2021common} and ROOT~\cite{vassilev2020automatic}. Gradients are used in objective-function optimizations, or in the propagation of uncertainty in different applications. %(from uncertainties in the parameters of tracks to the parameters of vertices or propagating the uncertainty in energy calibration to an invariant mass). 
A few projects aim to introduce AD-based pipelines for end-to-end sensitivity analysis in bigger scale systems such as reconstruction or down-stream analysis tasks \cite{de2019inferno,lukas2020}. The community creates events and documents to capture the growing interest~\cite{baydin2020differentiable}.

Recent advancements of production quality compilers like Clang allow tools to reuse the language parsing infrastructure, making it easier to implement source transformation AD. ADIC~\cite{bischof1997adic}, Enzyme~\cite{enzymeNeurips} and Clad~\cite{clad} are compiler-based AD tools using source transformation.
The increasing importance of AD is evident in newer programming languages such as Swift and Julia where it is integrated deep into the language~\cite{saeta2021swift, juliaDiff}.
Clad is a compiler-based source transformation AD tool for C++. Clad uses the high-level compiler program representation called \textit{abstract syntax tree} (AST). It can work as an extension to the Clang compiler, as an extension to the C++ interpreter Cling, or Clang-Repl, and is available in the ROOT software package. Clad implements forward and reverse accumulation modes and can provide higher order derivatives. It produces C++ code of the gradient, allowing for easy inspection and verification.

\section{Design and Implementation}

The Clang frontend builds a high-level program representation in the form of an AST. The programmatic AST synthesis might look challenging and laborious to implement, but yields several advantages. Having access to such a high-level code representation means being able to follow the high-level semantics of the algorithm and it facilitates domain-specific optimizations. This can be used to automatically generate code (re-targeting C++) for hardware acceleration with corresponding scheduling. Clang
supports an entire family of languages, such as C, C++, CUDA and OpenMP. For the most part, the AST for these languages is shared and thus the addition of CUDA support focuses only on the different constructs.
%Moreover, the AST is connected to the compiler diagnostics engine that allows it to produce precise and expressive diagnostics positioned at desired source locations.

The CUDA programming model provides a separation between the device (GPU) and host (CPU) using language attributes such as \textit{\_\_global\_\_}, \textit{\_\_host\_\_} and \textit{\_\_device\_\_}. A typical function that is to be differentiated can become a device executable function and its execution can be scheduled via kernels (functions that are marked with the \textit{\_\_global\_\_} attribute) for parallel computation. In order to maintain this support for the gradient function, it is sufficient for Clad to transfer the relevant attributes to the generated code and make the Clad-based run-time data structures compatible with CUDA by adding the relevant attributes.

\begingroup
\noindent
\begin{minipage}[h]{.51\textwidth}
\bigskip
\begin{cppcode*}{linenos=false}
#include "clad/Differentiator/Differentiator.h"
#define N 512
using arrtype = double[1];
// A device function to compute a gaussian
__device__ __host__ double gauss(double x,
    double p, double sigma) {
  double t = - (x - p) * (x - p) / 
               (2 * sigma * sigma);
  return pow(2 * M_PI, -1/2.0) * 
         pow(sigma, -0.5) * exp(t);
}
// Tells Clad to create a gradient
auto gauss_g = clad::gradient(gauss, "x, p");

// Forward declare the generated function so
// that the compiler can put it on the device.
void gauss_grad_0_1(double x, double p, 
    double sigma, clad::array_ref<double> _d_x, 
    clad::array_ref<double> _d_p)
    __attribute__((device)) 
    __attribute__((host));

\end{cppcode*}
\end{minipage}
\begin{minipage}[h]{.48\textwidth}
\bigskip
\begin{cppcode*}{linenos=false}
// The compute kernel including scheduling
__global__ void compute(double* x,
    double* p, double sigma, arrtype dx,
    arrtype dp) {
  int i = blockIdx.x * blockDim.x + 
          threadIdx.x;
  // Runs `N` different compute units,
  // each unit computes the gradient wrt
  // each parameter set
  if (i < N)
    gauss_grad_0_1(x[i], p[i], sigma, 
                   dx[i], dp[i]);
}

int main() {
  // CUDA device memory allocations...
  compute<<<N/256+1, 256>>>(dev_x, dev_p, 
                            dev_sigma,
                            dx, dp);
  // Utilize the results...
}
\end{cppcode*}
\end{minipage}
\captionof{listing}{Clad-CUDA support example of a function and generated gradient}\label{lst:cuda}
\endgroup
\vspace{0.5cm}
In essence, Clad operates on the same footing as the compiler’s template instantiation logic. As Clad \textit{visits} the AST, each node is cloned and differentiated. The differentiation rules are implemented in separate classes depending on the accumulation mode. In Listing~\ref{lst:cuda}, Clad differentiates a multidimensional normal distribution implemented as a device function. In this particular example, Clad generates the \textit{gauss\_grad\_0\_1} gradient executable in a CUDA environment. The generated gradient for the device function can be called from \textit{compute} in the same way the original function was, to take care of the level of parallelism.

Currently, preserving the parallel properties of the generated gradient relies on the user. In general, preserving parallel properties after AD transformation is still an ongoing research topic, however some interesting results can be achieved automatically based on the gradient content. We are working to automatically preserve the semantics of the generated code as in some cases the parallel reads from memory of the original function become parallel writes to memory in the produced gradient. In addition, the parallelism scheme of the original function might not be optimal for the gradient function. 
%because the gradient has more operations in general. 
We are working to make use of the information available in Clad to find a greater degree of parallelism for any produced code.
%The gradient obtained has full CUDA  support and its execution can thus be parallelised within a subsequent CUDA kernel.

%This approach to accelerating AD has several advantages: it works with legacy code; it adapts to multiple devices without modifying the computational part itself; the gradient code produced has good parallelism properties (e.g., a results array is produced for each parameter of the function); the differentiation is transparent to users, which means that users can easily inspect gradients and understand how a system evolves — useful in the context of reinforcement learning or meta learning; and by using C++ operators, we take full advantage of optimisation procedures on scheduling for multiple GPUs.

%In order to quantify the performance benefits this extension work is bringing forward, ongoing effort is dedicated to bench-marking procedures. We are working to compare Clad with other GPU aware AD tools and in this process to optimise Clad; one example of an optimization required is in the context of tape storage in looping conditions. 

\section{Integration and Results}

Interactive prototyping and exchanges for collaborative work is the backbone of research. Being a Clang plugin, Clad can easily integrate with the LLVM-based ecosystem. For example, Clad integrates with Jupyter notebooks, the C++ interactive interpreters Clang-Repl and Cling.

\begin{figure}[H]
  \centering
  \begin{subfigure}[b]{0.55\textwidth}
    \includegraphics[width=0.95\textwidth]{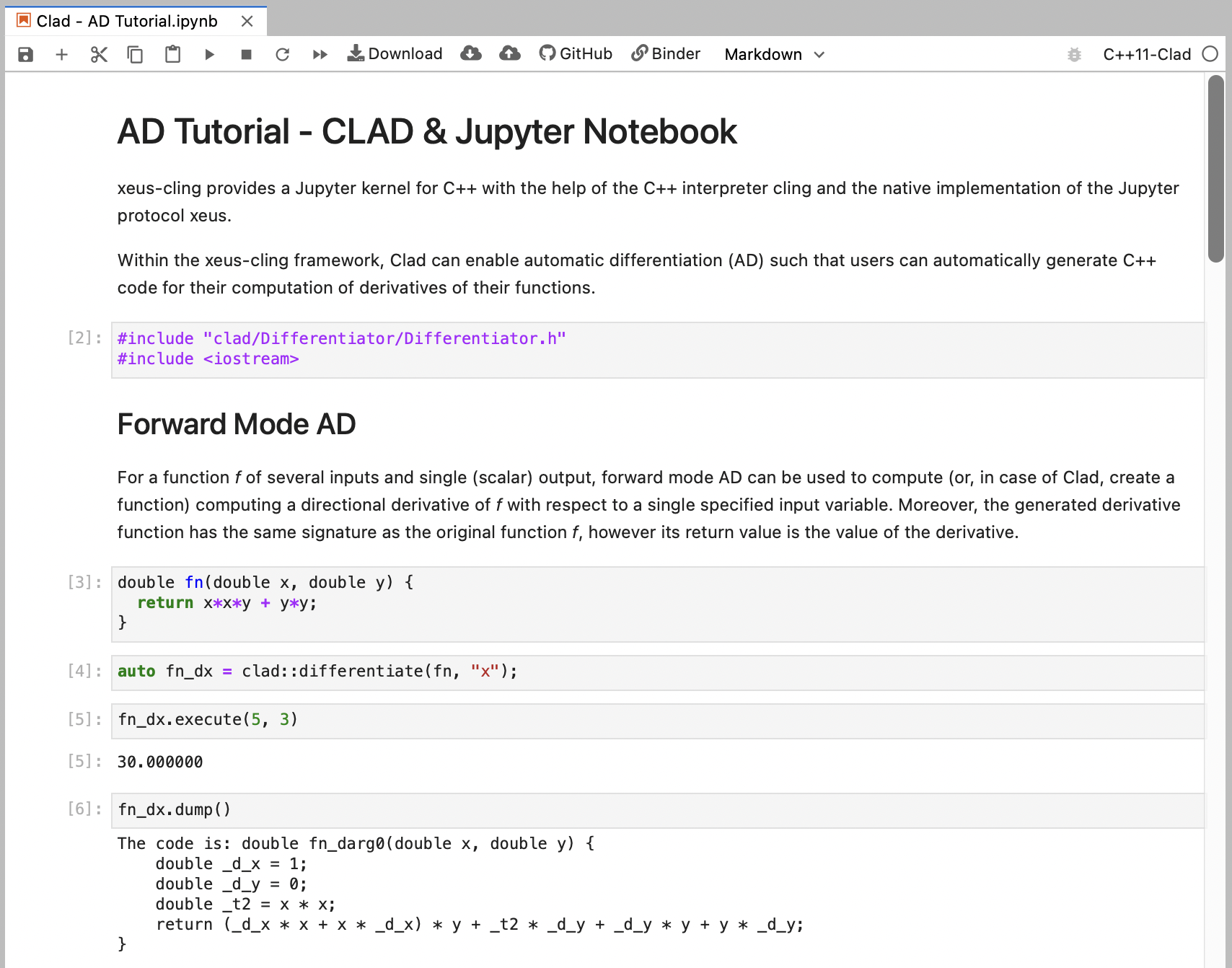}
    \caption{Jupyter} \label{fig:clad_interactive:jupyter}
  \end{subfigure}
  \hfill
   \begin{minipage}[b]{0.44\textwidth}
     \begin{subfigure}[b]{\linewidth}
        \includegraphics[width=0.95\textwidth]{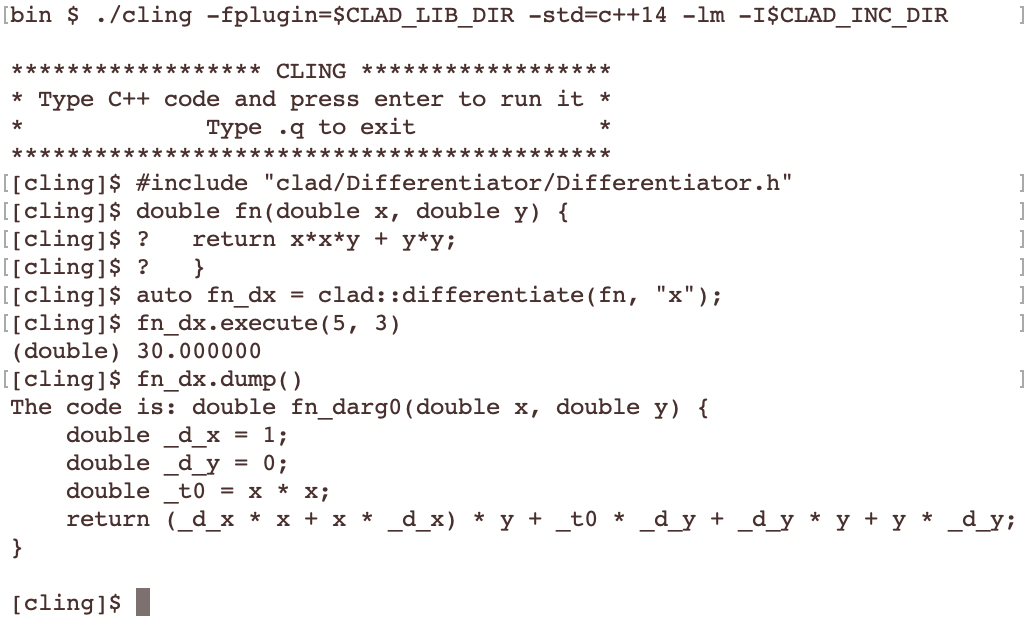}
        \caption{Cling}\label{fig:clad_interactive:cling}
        \vspace{0.4cm}
     \end{subfigure}
     \begin{subfigure}[b]{\linewidth}
       \includegraphics[width=0.95\textwidth]{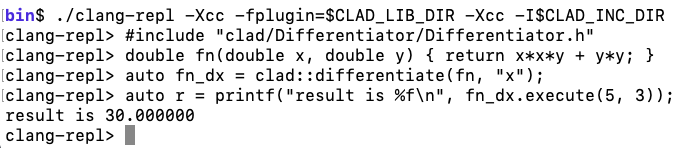}
       \caption{Clang-Repl} \label{fig:clad_interactive:clang-repl}
     \end{subfigure}
  \end{minipage}
  \caption{Clad Integrated in Interactive Environments}\label{fig:clad_interactive}
\end{figure}

Figure~\ref{fig:clad_interactive:jupyter} demonstrates that Clad can be used in a Jupyter notebook via xeus-cling. Xeus-cling enables interactive C++ for Jupyter notebooks.  Figure~\ref{fig:clad_interactive:cling} shows that Clad works with the interactive C++ interpreter Cling. Figure~\ref{fig:clad_interactive:clang-repl} demonstrates that Clad works in a terminal setup with Clang-Repl available in LLVM (since version 13). 

%\subsection{Clad in interactive environments}

%Firstly, Clad can be attached to Cling as a plugin in a similar way as it can be attached to Clang alongside. Cling is an interactive interpreter for C++ which is an implementation of the read-eval-print loop (REPL) concept also based on the LLVM infrastructure. Cling uses the Clang compiler as it provides incremental compilation. Thus, an user could prototype a CUDA dispatch of gradients quickly and easily by just attaching the two to their Cling environment. Secondly, Jupyter notebooks are omnipresent in today’s development process particularly in the data science space. xeus-cling [3] provides a Jupyter kernel for C++ with the help of the C++ interpreter Cling and the native implementation of the Jupyter protocol xeus. Thus, Clad can enable automatic differentiation (AD) such that users can automatically generate C++ code for their computation of derivatives of their functions in their Jupyter Notebook by simply having Clad installed alongside xeus-cling. A demonstrator of the two can be seen in Figure \ref{fig:clad_interactive}. 

%\begin{figure}[h]
%\centering
%\includegraphics[width=15cm, %height=5.5cm]{clad_interactive.pdf}
%\caption{Interactive Clad enabled AD}
%\label{fig:clad_interactive}
%\end{figure}

%\subsection{Clad for Data Analysis}

\begin{figure}[h]
\begin{subfigure}[b]{0.51\textwidth}
    \begin{cppcode*}{linenos=false}
auto f1 = new TF1("f1", "gaus");
auto h1 = new TH1D("h1", "h1", 1000, -5, 5);
double p1[] = {1, 0, 1.5}, p2[] = {100, 1, 3};
f1->SetParameters(p1); f1->SetParameters(p2);
h1->FillRandom("f1", 100000);
// Enable Clad in TFormula.
f1->GetFormula()->GenerateGradientPar();
auto r2 = h1->Fit(f1, "S G Q N"); // clad
// ...
\end{cppcode*}
\caption{Example use of Clad in TH1.}\label{fig:root_fitting}
\end{subfigure}
\begin{subfigure}[b]{0.49\textwidth}
  \includegraphics[width=\textwidth]{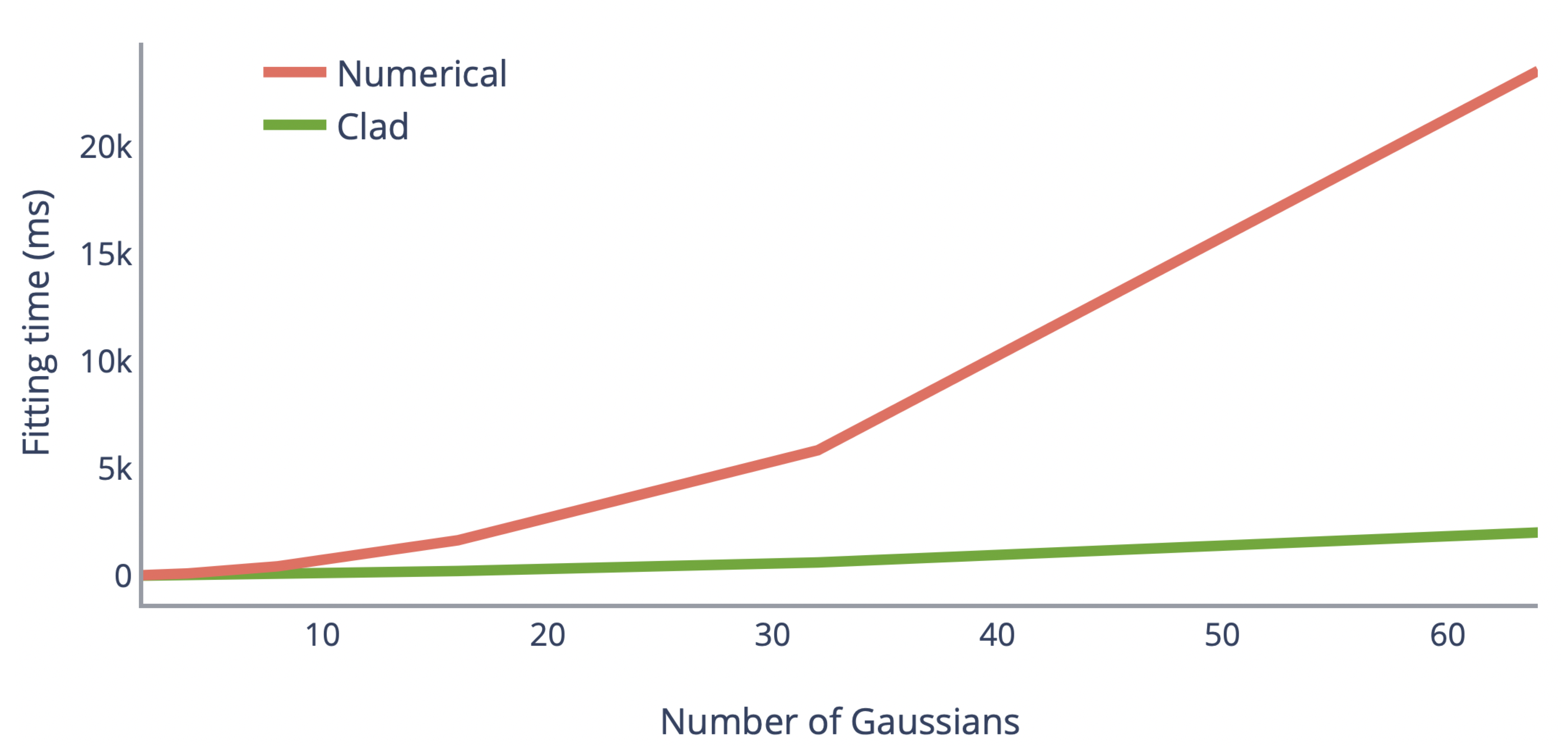}
  \caption{Scaling of many gaussian fits.}\label{fig:root_fitting_many}
\end{subfigure}
  \caption{Comparison of fitting time using gradients produced by Clad (in green) and numerically (red) of sums of Gaussians}
\end{figure}

AD with Clad is integrated in ROOT to provide an efficient way to calculate the gradient required for minimization or fitting with a function. This facility is available in ROOT's \textit{TFormula} class via the \textit{GradientPar} and the \textit{HessianPar} interfaces. The functionality can be used within the \textit{TF1} fitting interfaces as well, shown in figure~\ref{fig:root_fitting}. Replacing the numerical gradients yield promising results. In Figure~\ref{fig:root_fitting_many} we compare ROOT's implementation of numerical differentiation for generating gradients, and AD-produced gradients with Clad. As expected by the AD theory, gradients produced by reverse mode AD scale better as the time complexity of the computation is independent on the number of inputs. The better performance of Clad AD approach versus the numerical differentiation, via the central finite difference method, has been previously studied \cite{vassilev2020automatic}.
Further performance improvements can be achieved by moving the second order derivatives (the Hessian matrix) to use Clad. Currently, they are computed numerically as ROOT's minimizer must be adapted to use externally provided Hessians. 

We will work with the RooFit development team to integrate Clad into RooFit~\cite{Verkerke:2003ir}. RooFit  processes more computationally-intense operations and the benefit from automatically generated gradients will have a significant performance impact~\cite{bos2020faster}. Clad could be also used to differentiate the new batch-based CUDA backend in RooFit or re-target C++ gradients for GPU execution.

%It has replaced numerical gradient calculations for formula based functions. This means that Clad's reverse mode is used to compute the gradient of a given objective function (such as negative log-likelihood function). Clad functions are available in ROOT through the $Math/CladDerivator.h$ header. $TFormula$, a ROOT class that allows users to create and use multidimensional mathematical formulas, has tighter integration with Clad. Users can use the $GenerateGradientPar$ and $GenerateHessianPar$ functions to produce first order and second order derivatives, respectively, through Clad. $GradientPar$ and $HessianPar$ can then be used to evaluate the generated derivative at a specified point.

\section{Conclusion and Future Work}

%Software nowadays is built using networks of parameterised functional blocks which are then trained from examples using a form of gradient-based optimisation. The networks are defined procedurally in a data-dependent way (with loops and conditionals), which allows them to change dynamically as a function of the input data fed to them. Different from regular programming, differentiable programming is parameterised, automatically differentiated, and trainable/optimisable. Hence, the overall goal of differentiable programming is to quantify the susceptibility of the objective value at the output, using the chain rule to compute partial derivatives of the objective wrt each of the weights. In this way, the resulting algorithm can be interpreted as transforming the network evaluation function composed with the objective function under reverse mode AD (generalisation of the back-propagation procedure). The state-of-the-art results obtained by differentiable programming implementations have driven our motivation to add and extend CUDA support. 

In this paper, we introduced automatic differentiation and motivated its advantages for tasks oriented towards gradient-based optimization. We demonstrated the basic usage of Clad and described the advantages provided by it being a compiler-assisted tool. %The modification of the internal compiler data structures allowed more expressive diagnostics and good gradient debugging capabilities due to the fact that Clad can translate the generated gradient into high-level programming language.
We outlined the advancements in the area of CUDA in Clad. We demonstrated the ease of use via simple use cases of Clad and we showed how the tool integrates with other interactive systems such as Jupyter, Cling and Clang-Repl. We provided some performance results in using AD in ROOT's histogram fitting logic where we compared against the numerical differentiation approach.

We aim to further advance the GPU support of Clad, and to preserve better the parallel algorithm properties. One of the immediate next steps is to be able to differentiate the GPU kernel code automatically along with the already supported device function differentiation. 
The planned automatic kernel dispatch will utilize the information available in Clad to provide the optimal scheme for parallel computation for the produced gradient code. We plan to implement more advanced program analyses based on the Clang static analyzer infrastructure to allow more efficient code generation. We are working on extending the coverage of C++ and CUDA language features such as support of polymorphism (virtual functions) and differentiation with respect to aggregate types (for example class member variables). We work on a Clad-based backend for a statistical modelling tool with RooFit in order to provide efficient means to compute gradients.

%The intuition behind this is the assumption that in general, being parallel on the input array elements can lead to substantial speedup in terms of gradient execution. Thus, although there are generic solutions that can run on most use-cases involving large array data-sets where the data size and pattern are known, the dispatching solution can be further tuned for more general gradient computation able to achieve a substantial gain in performance. 

\section{Acknowledgments}

This project is supported by National Science Foundation under Grant OAC-1931408.
Some of the code contributions were facilitated by the 2021 Google Summer of Code program.

% \begin{table}[h]
% \caption{\label{ex}Table caption.}
% \begin{center}
% \begin{tabular}{llll}
% \br
% Head 1&Head 2&Head 3&Head 4\\
% \mr
% 1.1&1.2&1.3&1.4\\
% 2.1&2.2&2.3&2.4\\
% \br
% \end{tabular}
% \end{center}
% \end{table}

\printbibliography{}

\end{document}